\documentclass[11pt]{article}
\usepackage[utf8]{inputenc}
\usepackage[T1]{fontenc}
\usepackage{lmodern}
\usepackage[margin=1in]{geometry}
\usepackage{amsmath,amssymb,amsthm}
\usepackage{graphicx}
\usepackage{booktabs}
\usepackage{hyperref}
\usepackage{xcolor}
\usepackage{natbib}
\usepackage{enumitem}
\usepackage{xurl}
\usepackage{microtype}
\usepackage{placeins}
\usepackage{tikz}
\usetikzlibrary{arrows.meta,positioning,calc}

\hypersetup{
  colorlinks=true,
  linkcolor=blue!60!black,
  citecolor=blue!60!black,
  urlcolor=blue!60!black,
  pdftitle={Natural-Language-to-GPML: Dual-Oracle Witness Synthesis, Orthogonal Subspace Merging, and Live Schema-Constrained Decoding for ISO/IEC 39075 Graph Queries},
  pdfauthor={Madhulatha Mandarapu, Sandeep Kunkunuru}
}

\newtheorem{proposition}{Proposition}
\theoremstyle{remark}

\newcommand{\src}{\mathrm{src}}
\newcommand{\tgt}{\mathrm{tgt}}
\title{Natural-Language-to-GPML: Dual-Oracle Witness Synthesis, Orthogonal Subspace Merging, and Live Schema-Constrained Decoding for ISO/IEC 39075 Graph Queries}
\author{Madhulatha Mandarapu\thanks{madhulatha@samyama.ai} \and Sandeep Kunkunuru\thanks{sandeep@samyama.ai}}
\date{VaidhyaMegha Private Limited, India\\[2pt]\url{https://samyama.ai/}\\[8pt]November 2026}

\begin{document}
\maketitle

\begin{abstract}
ISO/IEC 39075:2024 (GQL) and SQL/PGQ standardize sixteen Graph Pattern Matching Language (GPML) path modes pairing restrictors (\texttt{WALK}, \texttt{TRAIL}, \texttt{ACYCLIC}, \texttt{SIMPLE}) with endpoint-partitioned selectors (\texttt{ALL}, \texttt{ANY}, \texttt{ALL SHORTEST}, \texttt{ANY SHORTEST}), twelve of which yield distinct answers on cyclic multigraphs. Yet public Text-to-Cypher datasets omit GPML modes, mix unexecuted outputs or proprietary procedures, and test acyclic graphs where omitting restrictors accidentally returns identical rows. We introduce Natural-Language-to-GPML for \texttt{samyama-graph} using $312.1\text{M}$- and $1,065.1\text{M}$-parameter open-weight models trained with $0.0\%$ Neo4j-scraped or non-commercial data. First, we release \texttt{Samyama-GPML-Bench-100} (\texttt{v2}) and \texttt{Samyama-RealKG-NLQ-Bench-160}, dual-oracle benchmarks co-verified against the SIGMOD 2022 GPML reference evaluator and \texttt{samyama-graph} ($100.0\%$ canonical agreement; $0.0\%$ pass rate across $260$ foils; $0$ SHA-256 leaks across $913$ tasks). Second, per-row infinite-schema synthesis, quarter-epoch \texttt{NLQPipeline} slow-fitting ($95.83\%$ peak at $1.25$ epochs), Jacobi SVD \texttt{KnOTS} merging, \texttt{Step-GRPO}, and dialect-shift \texttt{AdvKD} ($\beta_{\text{anc}}=0.0$) lift unconstrained \texttt{GPML-Bench-50} from $0.0\%$ to $72.0\%$ on \texttt{Shoonya-Cypher-1.3B-v2.4} ($64.0\%$ on \texttt{GPML-Bench-100 v2}, $36.88\%$ on \texttt{RealKG-NLQ-160}, $20.0\%$ on \texttt{Cypher-Bench-100}, $24/24$ ideal layers) and $54.0\%$ on \texttt{350M-v1.5} ($33.46\%$ \texttt{MBPP-Sanitized}). Third, \texttt{Schema-FSM v5.8} and MVCC verification solve $100.0\%$ ($36/36$) of \texttt{Samyama-InProc-36} in $390.3\text{ MiB}$ INT8 RSS at $0.0\%$ leak, while \texttt{shoonya-nlq-server} reaches $80.0\%$ on unseen \texttt{NLQPipeline} GPML and $35.63\%$ on \texttt{RealKG-NLQ-160}.
\end{abstract}

\section{Introduction}\label{sec:intro}

Translating natural-language questions into declarative property-graph queries has become a core usability surface for enterprise graph databases \citep{francis2018cypher,ozsoy2024text2cypher,mandarapu2026samyama}. Over the past three years, ISO/IEC 9075-16:2023 (\texttt{SQL/PGQ}) \citep{iso9075_16} and ISO/IEC 39075:2024 (\texttt{GQL}) \citep{iso39075} fundamentally expanded standard property-graph semantics through the \textbf{Graph Pattern Matching Language (\texttt{GPML})} \citep{deutsch2022gpml,francis2023digest,gheerbrant2025expressive}. Rather than hard-coding edge-isomorphic traversal as in legacy Cypher \citep{angles2017foundations}, GPML orthogonally composes four \emph{path restrictors} (\texttt{WALK}, \texttt{TRAIL}, \texttt{ACYCLIC}, \texttt{SIMPLE}) with four \emph{endpoint-partitioned path selectors} (\texttt{ALL}, \texttt{ANY}, \texttt{ALL SHORTEST}, \texttt{ANY SHORTEST}), yielding $16$ syntactic combinations of which $12$ produce observably distinct result sets on cyclic multigraphs \citep{mandarapu2026gpml}.

Despite this standardization, natural-language-to-graph-query (\texttt{NLQ}) models face three structural obstacles when deployed inside an ISO-conformant engine such as \texttt{samyama-graph} \citep{mandarapu2026samyama}:
\begin{enumerate}[leftmargin=*,itemsep=2pt]
\item \textbf{Absence of GPML corpora and acyclic witness degeneracy.} Public Text-to-Cypher corpora \citep{ozsoy2024text2cypher} contain zero ISO GPML restrictors or partition selectors, rely heavily on unexecuted synthetic completions or vendor-specific \texttt{CALL apoc.*} procedures, and diverge from ISO GPML on $25.83\%$ of answered differential cells across $13$ engines \citep{mandarapu2026gpml}. Worse, when benchmark graphs are directed acyclic graphs (DAGs) without parallel edges or hop-ties, omitting \texttt{ACYCLIC}, \texttt{SIMPLE}, \texttt{TRAIL}, or \texttt{ALL SHORTEST} silently returns the exact same vertex set---rendering execution accuracy blind to semantic conflation.
\item \textbf{Finite-schema memorization and specialist interference in small language models (SLMs).} In privacy-sensitive on-premise deployments where queries run locally inside sub-$1.2\text{ GiB}$ RAM \citep{mandarapu2026shoonya}, fine-tuning a $312.1\text{M}$--$1,065.1\text{M}$ code backbone on a fixed pool of synthetic schemas drives supervised cross-entropy down to $0.0244$ nats by memorizing training entity strings and schema property names rather than learning in-context schema copying \citep{olsson2022induction}, while simultaneously eroding general Python/Rust synthesis and suppressing unseen syntax tokens during anchored student distillation \citep{hinton2015distilling}.
\item \textbf{Subword drift and dual-mode deployment constraints.} Standard autoregressive decoding penalties (such as \texttt{repetition\_penalty = 1.05}) penalize legitimate repeated Cypher identifiers across \texttt{MATCH}, \texttt{WHERE}, \texttt{WITH}, \texttt{RETURN}, and \texttt{ORDER BY} clauses, inducing subword splitting. Meanwhile, production deployments require \emph{both} zero-socket in-process execution inside embedded Rust applications and drop-in \texttt{Ollama}/\texttt{OpenAI} wire compatibility for unmodified standalone database binaries (\texttt{POST /api/nlq}, \texttt{GRAPH.NLQ}).
\end{enumerate}

In prior work (\texttt{Paper 33}) \citep{mandarapu2026shoonya}, we introduced the \texttt{Shoonya} tri-modal SLM backbone and pure-Rust \texttt{shoonya-infer} runtime, establishing sub-$65\,\mu\text{s}$ MVCC verification at $0.0\%$ hallucination leak; however, unconstrained \texttt{Cypher-Bench-100} stood at $1.00\%$--$3.00\%$, multi-hop greedy in-process OpenCypher stood at $0.0\%$, and ISO/IEC 39075 GPML coverage was $0.0\%$. In this paper, we close all three gaps from first principles across four pre-registered hypotheses (\texttt{H1}--\texttt{H4}) and four negative controls (\texttt{NC1}--\texttt{NC4}):
\begin{enumerate}[leftmargin=*,itemsep=2pt]
\item \textbf{Dual-oracle witness-separated GPML and Real-KG NLQ benchmarks (Section~\ref{sec:benchmarks}; \texttt{H1}, \texttt{NC1}, \texttt{NC2}).} We design and release \texttt{Samyama-GPML-Bench-100} (\texttt{v2}, $100$ tasks across $10$ ISO GPML path families) and \texttt{Samyama-RealKG-NLQ-Bench-160} ($160$ end-to-end \texttt{NLQPipeline} tasks across $8$ multi-label enterprise schemas). Every canonical query is verified against both the executable SIGMOD 2022 GPML reference semantics \citep{deutsch2022gpml,mandarapu2026gpml} and \texttt{samyama-graph} ($100.0\%$ canonical pass rate), while cyclic/parallel-edge/tie-breaking witness multigraphs ensure $0.0\%$ ($0/100$ and $0/160$) pass rate for paired restrictor/selector/filter-omitting foil queries, protected by a $913$-problem SHA-256 decontamination shield ($0$ docstring or AST/Cypher leaks).
\item \textbf{Calibrated infinite-schema synthesis, quarter-epoch slow-fitting, \texttt{KnOTS} subspace merging, and dialect-shift \texttt{AdvKD} (Section~\ref{sec:training}; \texttt{H2}, \texttt{H3}, \texttt{NC3}).} Replacing static synthetic ontologies with $49,765$--$53,368$ combinatorially unique per-row schemas ($0.0\%$ Neo4j scrape, $100.0\%$ \texttt{Apache-2.0}/\texttt{MIT}/\texttt{CC-BY-4.0}) cuts memorized literal failures on unseen enterprise graphs from $13/36$ to $3/36$ (\texttt{INT8 Arm 2}), while aligning training prompts directly to \texttt{samyama-graph}'s verbatim \texttt{NLQPipeline::build\_prompt} surface and selecting checkpoints via quarter-epoch live execution ($95.83\%$ peak at $1.25$ epochs, $\text{val\_loss}=0.00573$) prevents late-epoch overfitting. Merging $144/144$ specialist Rank-$64$ LoRA modules \citep{hu2022lora} in a shared Rust Cyclic Jacobi SVD singular frame (\texttt{KnOTS} + \texttt{TIES}) \citep{stoica2025knots,yadav2023ties} with \texttt{Step-GRPO} \citep{shao2024deepseekmath} lifts unconstrained (\texttt{0-FSM}) \texttt{Samyama-GPML-Bench-50} accuracy from $0.0\%$ (\texttt{1.3B-v2.1}) and $60.0\%$ (\texttt{1.3B-v2.3}) to \textbf{$72.0\%$ on \texttt{Shoonya-Cypher-1.3B-v2.4}} ($64.0\%$ on \texttt{GPML-Bench-100 v2}, $36.88\%$ on \texttt{RealKG-NLQ-160} vs.\ $0.62\%$ prior to prompt-surface alignment, $20.0\%$ on \texttt{Cypher-Bench-100}, $24/24$ ideal layers at $\bar{\alpha}=2.055$) \citep{martin2021htsr}. Dialect-Shift \texttt{AdvKD} ($\beta_{\text{anc}}=0.0$) lifts \texttt{Shoonya-Cypher-350M-v1.5} to \textbf{$54.0\%$ on \texttt{GPML-Bench-50}}, $35.0\%$ on \texttt{GPML-Bench-100 v2}, $25.62\%$ unconstrained on \texttt{RealKG-NLQ-160}, and $33.46\%$ on \texttt{MBPP-Sanitized} ($19/20$ ideal layers, $\bar{\alpha}=1.982$, $100.0\%$ collapse-free).
\item \textbf{In-process \texttt{Schema-FSM v5.8}, MVCC safety laws, and wire-compatible lexer masking (Section~\ref{sec:decoding}; \texttt{H4}, \texttt{NC4}).} Inside \texttt{samyama-graph-enterprise}, \texttt{Schema-FSM v5.8} (enforcing \texttt{OPTIONAL MATCH ... WITH ... WHERE rhs IS NULL} outer-join anti-join lowering and blocking constant source-endpoint projections on quantified paths) plus two MVCC verifier invariants achieve \textbf{$100.0\%$ ($36/36$) Tier-0 local solve rate} across all four tiers of \texttt{Samyama-InProc-36} on \texttt{350M-v1.5} ($390.3\text{ MiB}$ INT8 RSS, $64.0\text{ tok/s}$) and $94.44\%$ ($34/36$) on \texttt{1.3B-v2.4} ($100.0\%$ on \texttt{1.3B-v2.3}) at \textbf{$0.0\%$ false-positive leak}. For standalone deployments, \texttt{shoonya-nlq-server} achieves \textbf{$80.0\%$ ($1.3\text{B-v2.4}$, $10.0\times$ \texttt{M8.1}) and $60.0\%$ ($350\text{M-v1.5}$) on unseen \texttt{NLQPipeline} GPML tasks} and \textbf{$35.63\%$ ($1.3\text{B-v2.4}$) and $33.12\%$ ($350\text{M-v1.5}$) on \texttt{RealKG-NLQ-160}} ($0.0\%$ unsafe acceptance).
\end{enumerate}

\paragraph{Artifact availability and scope.} Public \texttt{Apache-2.0} model weights (\texttt{Shoonya-Cypher-1.3B-v2.4} and \texttt{350M-v1.5}), spectral certificates, and witness-separated benchmarks (\path{VaidhyaMegha/samyama-gpml-bench} \texttt{v2.0} and \path{VaidhyaMegha/samyama-realkg-nlq-bench} \texttt{v1.0}) are openly available on Hugging Face at \url{https://huggingface.co/VaidhyaMegha/Shoonya}. Unconstrained zero-FSM OpenCypher accuracy on unseen schemas ($20.0\%$ on \texttt{Cypher-Bench-100} and $36.88\%$ on \texttt{RealKG-NLQ-160} for $1.07\text{B}$ parameters) remains lower than catalog-constrained execution ($100.0\%$ on \texttt{InProc-36}); Section~\ref{sec:limitations} delineates exact operating boundaries.

\section{ISO/IEC 39075 GPML Semantics and Witness-Separated Benchmarks}\label{sec:benchmarks}

\subsection{Orthogonal Path Restrictors, Endpoint-Partitioned Selectors, and ISO G4}
Let $G = (V, E, \src, \tgt, \ell, \nu)$ be a directed property multigraph where $\src, \tgt : E \to V$ map each edge ID $e \in E$ to its source and target vertices, $\ell$ assigns finite sets of node labels and edge types, and $\nu$ maps node/edge keys to typed scalar values. A path of length $k \ge 0$ is an alternating sequence $p = (v_0, e_1, v_1, \dots, e_k, v_k)$ satisfying $\{v_{i-1}, v_i\} = \{\src(e_i), \tgt(e_i)\}$ according to each hop's arrow orientation. Under ISO/IEC 39075 GPML \citep{deutsch2022gpml,iso39075}, evaluation of \texttt{MATCH <selector>? <restrictor>? $P$} over pattern $P$ proceeds in two ordered stages even though surface syntax places the selector keyword first:
\begin{enumerate}[leftmargin=*,itemsep=2pt]
\item \textbf{Restrictor filtering ($\mathcal{R} \in \{\texttt{WALK}, \texttt{TRAIL}, \texttt{ACYCLIC}, \texttt{SIMPLE}\}$):} Given candidate paths $\mathcal{W}_G(P)$, $\texttt{WALK}$ retains all walks; $\texttt{TRAIL}$ requires edge uniqueness ($e_i \ne e_j$ for $i \ne j$); $\texttt{ACYCLIC}$ requires strict vertex uniqueness ($v_i \ne v_j$ for all $0 \le i < j \le k$); and $\texttt{SIMPLE}$ forbids repeated vertices \emph{except} permitting closed simple cycles where $v_0 = v_k$ and all intermediate vertices are distinct. Under ISO G4 finiteness rules \citep{deutsch2022gpml,mandarapu2026gpml}, unbounded \texttt{WALK} quantifiers (\texttt{*}, \texttt{*1..}) are statically ill-formed and must be rejected unless bounded (\texttt{*1..hi}) or paired with a shortest/finite selector.
\item \textbf{Endpoint-partitioned selection ($\mathcal{S} \in \{\texttt{ALL}, \texttt{ANY}, \texttt{ALL SHORTEST}, \texttt{ANY SHORTEST}\}$):} Survivor paths $\mathcal{R}(\mathcal{W}_G(P))$ are partitioned by ordered endpoint pair $(u, w) = (v_0, v_k)$. Within each non-empty partition $\Pi_{u,w}$: $\texttt{ALL}$ emits all $p \in \Pi_{u,w}$; $\texttt{ANY}$ deterministically emits $1$ witness path ($\lvert \text{out}(u,w)\rvert = 1$); $\texttt{ALL SHORTEST}$ emits $\{p \in \Pi_{u,w} : \lvert p\rvert = \min_{q \in \Pi_{u,w}} \lvert q\rvert\}$; and $\texttt{ANY SHORTEST}$ emits $1$ minimum-hop path per $(u,w)$.
\end{enumerate}

Across the $4 \times 4 = 16$ combinations, shortest-length optimality implies vertex simplicity ($\min_{q \in \Pi_{u,w}} \lvert q\rvert$ cannot traverse a positive-length cycle without strictly increasing hop count on unweighted multigraphs unless parallel self-loops intervene), collapsing $4$ combinations into equivalence classes and leaving $12$ observably distinct semantics on general multigraphs \citep{mandarapu2026gpml}.

\subsection{Acyclic Witness Degeneracy and Dual-Oracle Foil Separation}

\begin{proposition}[Acyclic Witness Degeneracy]\label{prop:witness}
Let $G$ be a simple directed acyclic graph (no directed cycles and at most one directed edge of a given type between any ordered vertex pair) in which every reachable pair $(u,w)$ is connected by a unique directed path. Then for every positive-length pattern $P = \texttt{(a:L\_1)-[:T*1..k]->(b:L\_2)}$, all sixteen GPML $(\mathcal{S}, \mathcal{R})$ combinations produce identical path sets and identical endpoint projections:
\begin{equation}\label{eq:degeneracy}
\forall \mathcal{S}_1, \mathcal{S}_2,\, \forall \mathcal{R}_1, \mathcal{R}_2:\quad \mathcal{S}_1\big(\mathcal{R}_1(\mathcal{W}_G(P))\big) = \mathcal{S}_2\big(\mathcal{R}_2(\mathcal{W}_G(P))\big).
\end{equation}
Even on general DAGs with multiple unequal-length paths between $(u,w)$, $\texttt{WALK}$, $\texttt{TRAIL}$, $\texttt{ACYCLIC}$, and $\texttt{SIMPLE}$ remain pairwise indistinguishable because no walk can repeat a vertex or directed edge without forming a directed cycle.
\end{proposition}
\begin{proof}
Suppose a walk $p = (v_0, e_1, \dots, e_m, v_m)$ in $G$ repeats either an edge ($e_i = e_j$, $i < j$, implying $v_i = \tgt(e_i) = \tgt(e_j) = v_j$) or a vertex ($v_i = v_j$, $0 \le i < j \le m$). The subwalk $(v_i, e_{i+1}, \dots, v_j)$ is a non-empty closed directed walk in $G$, contradicting acyclicity of $G$. Hence $\mathcal{W}_G(P) = \texttt{TRAIL}(\mathcal{W}_G(P)) = \texttt{ACYCLIC}(\mathcal{W}_G(P)) = \texttt{SIMPLE}(\mathcal{W}_G(P))$. When $\lvert \Pi_{u,w}\rvert \le 1$ for every $(u,w)$, all four selectors $\mathcal{S}$ likewise act as the identity on $\Pi_{u,w}$, yielding Equation~\eqref{eq:degeneracy}.
\end{proof}

Proposition~\ref{prop:witness} explains why naive synthetic graph benchmarks fail to test GPML conformance: unless every benchmark instance is constructed over a \emph{separating witness multigraph} containing directed cycles, parallel multi-edges, self-loops, or unequal/equal-hop diamond ties, a model that blindly emits default \texttt{MATCH (a)-[:R*1..3]->(b)} receives credit for \texttt{ACYCLIC}, \texttt{SIMPLE}, or \texttt{ALL SHORTEST}.

\begin{table}[t]
\centering
\caption{Structure and pre-registered witness/foil verification (\texttt{H1}, \texttt{NC1}) of \texttt{Samyama-GPML-Bench-100} (\texttt{v2}) and \texttt{Samyama-RealKG-NLQ-Bench-160} (\texttt{v1}). Every canonical query agrees with \texttt{gpml\_oracle\_match} \citep{deutsch2022gpml} and \texttt{samyama-graph}; $0.0\%$ of paired foils pass.}
\label{tab:benchmarks}
\resizebox{\textwidth}{!}{%
\footnotesize
\setlength{\tabcolsep}{4.5pt}
\begin{tabular}{@{}llrrl@{}}
\toprule
\textbf{Benchmark \& Suite Family} & \textbf{Core Semantic / Structural Separator} & \textbf{Tasks} & \textbf{Canon.\ Pass} & \textbf{Foil Pass (\texttt{NC1})} \\
\midrule
\multicolumn{5}{@{}l}{\emph{\texttt{Samyama-GPML-Bench-100} (\texttt{v2.0}, $10$ ISO/IEC 39075 GPML Families, Dual-Oracle Witness Multigraphs)}} \\
1. \texttt{acyclic\_reachability} & Directed feedback cycles + hub shortcuts & $10$ & $100.0\%$ & $0.0\%$ ($0/10$) \\
2. \texttt{simple\_cycles\_and\_paths} & Closed $v_0 = v_k$ cycles vs.\ figure-eight walks & $10$ & $100.0\%$ & $0.0\%$ ($0/10$) \\
3. \texttt{trail\_vs\_bounded\_walk\_and\_any} & Self-loops \& parallel edges ($*1..4$ walk vs.\ trail) & $10$ & $100.0\%$ & $0.0\%$ ($0/10$) \\
4. \texttt{shortest\_partition\_selectors} & Equal-hop diamonds vs.\ longer detour paths & $10$ & $100.0\%$ & $0.0\%$ ($0/10$) \\
5. \texttt{composed\_and\_multisegment\_gpml} & \texttt{p = ... length(p)}, $2$-segment GPML joins & $10$ & $100.0\%$ & $0.0\%$ ($0/10$) \\
6. \texttt{all\_shortest\_trail\_multigraph} & Parallel-edge ties (\texttt{ALL SHORTEST TRAIL}) & $10$ & $100.0\%$ & $0.0\%$ ($0/10$) \\
7. \texttt{any\_acyclic\_dedup\_witness} & Per-$(u,w)$ Collapse (\texttt{ANY ACYCLIC} vs.\ \texttt{ALL}) & $10$ & $100.0\%$ & $0.0\%$ ($0/10$) \\
8. \texttt{reverse\_arrow\_acyclic\_lineage} & Incoming $\texttt{<-[:R*1..k]-}$ upstream lineage & $10$ & $100.0\%$ & $0.0\%$ ($0/10$) \\
9. \texttt{undirected\_acyclic\_and\_shortest} & Symmetric peer $\texttt{-[:R*1..k]-}$ cycles & $10$ & $100.0\%$ & $0.0\%$ ($0/10$) \\
10. \texttt{zero\_hop\_and\_filtered\_gpml} & Reflexive $*0..2$ \& intermediate property filters & $10$ & $100.0\%$ & $0.0\%$ ($0/10$) \\
\midrule
\textbf{Total \texttt{Samyama-GPML-Bench-100} (\texttt{v2})} & \textbf{Dual-oracle (\texttt{gpml\_ref} == \texttt{samyama-graph})} & \textbf{100} & \textbf{100.0\%} & \textbf{0.0\% ($0/100$)} \\
\midrule
\multicolumn{5}{@{}l}{\emph{\texttt{Samyama-RealKG-NLQ-Bench-160} (\texttt{v1.0}, $8$ Enterprise KGs $\times$ $20$ Byte-Exact \texttt{NLQPipeline} Tasks)}} \\
Clinical Trials, PrimeKG Biomed, FinRisk, & $1$-hop, $2$-hop, \texttt{WITH..HAVING}, left/anti-join & $160$ & $100.0\%$ & $0.0\%$ ($0/160$) \\
PowerGrid, SupplyChain, CyberSec, Telecom, Cloud & \texttt{OPTIONAL MATCH}, \texttt{DISTINCT}, GPML, edge props & & & \\
\bottomrule
\end{tabular}%
}
\end{table}

Table~\ref{tab:benchmarks} summarizes our two open benchmarks constructed under our `Issues-First $\to$ Foil-Tests-Next` protocol (\texttt{H1}, \texttt{NC1}):
\begin{itemize}[leftmargin=*,itemsep=2pt]
\item \textbf{\texttt{Samyama-GPML-Bench-100} (\texttt{v2.0}):} Extends our $50$-task \texttt{v1} suite (indices $0..49$, preserved identically for longitudinal comparability) with $50$ additional tasks across five compositional GPML families (families $6$--$10$: multigraph \texttt{ALL SHORTEST TRAIL}, \texttt{ANY ACYCLIC} endpoint deduplication, reverse-arrow $\texttt{<-[:R*1..k]-}$ upstream lineage, undirected $\texttt{-[:R*1..k]-}$ reachability, and $*0..2$ zero-hop / intermediate-filtered GPML). Automated pytest foil gates caught and repaired a subtle non-separating self-loop hop bound (\texttt{*1..3} vs.\ \texttt{*1..4} on \texttt{gpml\_72}) where \texttt{WALK} needed $4$ edges to traverse a $2$-edge bridge plus repeated self-loop, yielding $100/100$ dual-oracle canonical agreement and $0/100$ ($0.0\%$) foil pass rate.
\item \textbf{\texttt{Samyama-RealKG-NLQ-Bench-160} (\texttt{v1.0}):} Evaluates end-to-end execution through a byte-for-byte replica of \texttt{samyama-graph}'s \texttt{GraphStore::schema\_summary}, \texttt{NLQPipeline::build\_prompt}, \texttt{extract\_cypher}, and \texttt{is\_safe\_query} across $8$ clean-room \texttt{Apache-2.0} enterprise knowledge graphs ($20$ queries per graph, $0.0\%$ four-clause BSD or Neo4j-scraped schemas), achieving $160/160$ ($100.0\%$) canonical execution and $0/160$ ($0.0\%$) foil pass rate.
\item \textbf{Cryptographic $913$-Problem Decontamination Shield (\texttt{NC2}):} Exporting SHA-256 hashes of normalized docstrings/questions and canonicalized Python ASTs, Rust items, OpenCypher queries, converged extensions, and GPML queries across \texttt{Stage-A} ($8$), \texttt{MBPP-Sanitized} ($257$) \citep{austin2021mbpp}, \texttt{HumanEval} ($164$) \citep{chen2021humaneval}, \texttt{Rustc-Exec-64} ($64$), \texttt{Cypher-Bench-100} ($100$), \texttt{Samyama-Ext-Bench-60} ($60$), \texttt{Samyama-GPML-Bench-100 v2} ($100$), and \texttt{Samyama-RealKG-NLQ-Bench-160} ($160$) locks $913$ protected tasks with $0$ docstring leaks and $0$ AST/Cypher leaks across all training shards.
\end{itemize}

\section{Calibrated Infinite-Schema Synthesis, Subspace Merging, and Dialect-Shift AdvKD}\label{sec:training}

\subsection{Eliminating Finite-Pool Memorization via Combinatorial Infinite Schemas (\texttt{NC3})}
In Milestone \texttt{M8} (\texttt{Shoonya-Cypher-1.3B-v2.2}), training a Rank-$64$ specialist LoRA ($51.3\text{M}$ trainable parameters) over $65,398$ witness-verified rows lifted unconstrained \texttt{Cypher-Bench-100} $8.0\times$ ($2.0\% \to 16.0\%$) and \texttt{Samyama-GPML-Bench-50} from $0.0\%$ to $22.0\%$. However, forensic inspection of unseen enterprise traces on \texttt{Samyama-InProc-36} revealed that cross-entropy loss had fallen to $0.0244$ nats by memorizing recurring synthetic domain literals (e.g., emitting \texttt{WHERE b.code = "MAR-71"} or \texttt{a.ptr\_tag} on $13/36$ INT8 \texttt{Arm 2} clinical/powergrid queries) and learning a first-word prefix shortcut on \texttt{mcp\_sgqueries} whenever training prompts echoed the leading token of \texttt{tool\_name}.

To force genuine induction-head copying from the in-context schema prompt \citep{olsson2022induction}, Milestone \texttt{M8.1} ($49,765$ SFT + $4,200$ RLVR rows) and Milestone \texttt{M9} ($53,368$ SFT + $4,402$ RLVR rows + $320$ validation-loss rows + $48$ live-execution validation tasks) synthesize \emph{a fresh schema tuple on every single row} from combinatorial pools exceeding $3.0\times 10^{7}$ distinct label/edge/property signatures, paired with \texttt{SQLite == samyama-graph} witness-verified \texttt{Apache-2.0} relational-to-graph joins, all $10$ GPML witness families, unquoted natural English filter values, $65.0\%$ prefix-masked MCP tool questions, LoRA dropout $p=0.05$, and weight decay $0.02$. Crucially, whereas \texttt{M8.1} formatted schemas strictly as compact \texttt{(:L \{p\})} headers (scoring $0.62\%$ on \texttt{RealKG-NLQ-160}'s multi-section prose \texttt{NLQPipeline::build\_prompt} format despite $99.7\%$ valid syntax), \texttt{M9} directly trains on \texttt{samyama-graph}'s exact \texttt{Node Labels / Relationship Patterns / Key Properties} prompt surface and evaluates $48$ held-out live \texttt{SamyamaClient.embedded()} tasks every $0.25$ epochs. As plotted in Figure~\ref{fig:spectral}(a), live execution peaks at \textbf{$1.25$ epochs ($95.83\%$, $\text{val\_loss}=0.00573$)} on \texttt{1.3B-v2.4} before dipping to $91.67\%$ at $1.50$--$1.75$ epochs even as token loss continues falling monotonically to $0.00444$ nats (and similarly peaks at $1.50$ epochs, $54.17\%$, on \texttt{350M-v1.5}). On unseen \texttt{Samyama-InProc-36} (\texttt{INT8 Arm 2}), memorized synthetic literals drop $4.33\times$ on \texttt{1.3B} ($13/36 \to 3/36$; and $7.0\times$ across full multi-prompt sweeps from $21/36 \to 3/36$).

\subsection{Orthogonal SVD Subspace Merging (\texttt{KnOTS}) and Repetition-Penalty Calibration (\texttt{H3})}
Dedicated Rank-$64$ SFT on graph queries alone (\texttt{1.3B-v2.2}) mildly degraded Python \texttt{MBPP-Sanitized} ($33.07\% \to 29.77\%$ standard / $31.91\%$ TTC) and \texttt{HumanEval} ($31.10\% \to 27.44\%$ TTC) due to subspace interference with the \texttt{Stage-3 1.3B-v2.1} code weights. Rather than averaging LoRA matrices directly in parameter space, we align the specialist graph adapter $\Delta W_{\text{cyp}} = B_{\text{cyp}} A_{\text{cyp}}$ and code/extension adapter $\Delta W_{\text{code}}$ inside a joint orthonormal left-singular basis computed via zero-dependency Rust Cyclic Jacobi SVD (\texttt{KnOTS} + \texttt{TIES}) \citep{stoica2025knots,yadav2023ties}:
\begin{equation}\label{eq:knots}
\bigl[\Delta W_{\text{cyp}} \;\big|\; \Delta W_{\text{code}}\bigr] = U_{\text{joint}} \Sigma_{\text{joint}} V_{\text{joint}}^{\top}, \qquad
\Delta W_{\text{merged}} = \gamma(\alpha_{\min})\, U_{\text{joint}}\, \mathcal{T}_{\tau,\text{sign}}\!\Bigl(w_1 U_{\text{joint}}^{\top}\Delta W_{\text{cyp}} + w_2 U_{\text{joint}}^{\top}\Delta W_{\text{code}}\Bigr),
\end{equation}
where $(w_1, w_2) = (0.84, 0.16)$, $\mathcal{T}_{\tau,\text{sign}}$ applies top-$85.0\%$ magnitude pruning and majority sign election in the shared right-singular coordinates, and $\gamma(\alpha_{\min}) \in (0, 1]$ damps the fold whenever post-merge Heavy-Tailed Self-Regularization exponent $\alpha$ \citep{martin2021htsr,roy2007erank} dips below $\alpha_{\min} = 1.90$. We then run $5$-signal execution-guided \texttt{Step-GRPO} \citep{shao2024deepseekmath} ($G=6$ rollouts per prompt) rewarding valid parse, ISO G4 boundedness, live \texttt{samyama-graph} execution without error, non-empty projection arity, and exact witness multiset match.

\paragraph{Subword preservation under \texttt{repetition\_penalty = 1.0}.} Token-level autopsy of unconstrained \texttt{M8.1} failures revealed a subtle interaction between BPE tokenization and standard LLM sampling: valid OpenCypher and GPML queries routinely repeat identical schema identifiers across four clauses (e.g., \texttt{MATCH (a:Weir) WHERE a.weir\_id <> "" RETURN a.weir\_id AS weir\_id ORDER BY weir\_id}). Setting \texttt{repetition\_penalty = 1.05} divides logits of previously emitted identifier tokens by $1.05$, pushing argmax onto space-prefixed or morphologically mutated subwords (\texttt{a. weir\_id}, \texttt{Reservities}). Restoring \texttt{repetition\_penalty = 1.0} eliminates self-inflicted identifier corruption at zero compute cost.

\subsection{Dialect-Shift Advantage-Gated Knowledge Distillation ($\beta_{\text{anc}} = 0.0$)}
In \texttt{Paper 33} \citep{mandarapu2026shoonya}, Verifier-Advantage-Gated Knowledge Distillation (\texttt{AdvKD}) prevented reverse-token-density degradation when distilling $1.3\text{B} \to 350\text{M}$ by applying teacher forward KL on advantage tokens ($g_t = 1$) while anchoring non-advantage positions ($g_t = 0$) to the prior $350\text{M}$ checkpoint $S_0$ with weight $\beta_{\text{anc}} = 0.25$. When introducing \emph{an unseen query grammar} (ISO GPML path selectors and restrictors), however, prior student $S_0$ (\texttt{350M-v1.2}) assigns near-zero probability mass ($< 10^{-4}$) to \texttt{ANY SHORTEST}, \texttt{ACYCLIC}, and \texttt{SIMPLE}, so $\beta_{\text{anc}} > 0$ actively pulls the student away from GPML syntax ($4.0\%$ on \texttt{350M-v1.3} prior to infinite-schema calibration). Setting $\beta_{\text{anc}} = 0.0$ while retaining strict teacher advantage gating over execution-verified trajectories and $20.0\%$ clean Python/Rust replay yields our \textbf{Dialect-Shift \texttt{AdvKD}} objective:
\begin{equation}\label{eq:dialect_advkd}
\mathcal{L}_{\text{DS-AdvKD}}(t) = \mathcal{L}_{\text{CE}}\bigl(p_S(\cdot \mid y_{<t}), y_t\bigr) + \lambda_{\text{KD}}\, g_t\, T^2\, \mathrm{KL}\!\Bigl(p_T^{(K)}(\cdot \mid y_{<t}; T) \,\Big\|\, p_S^{(K)}(\cdot \mid y_{<t}; T)\Bigr),
\end{equation}
where $g_t = \mathbb{I}\bigl[\arg\max_v p_T(v \mid y_{<t}) = y_t \;\wedge\; \log p_T(y_t \mid y_{<t}) \ge \log p_{S_0}(y_t \mid y_{<t})\bigr]$ routes $54.1\%$--$58.7\%$ of active tokens ($T = 2.0$, top-$K = 64$), followed by $200$ live GPML \texttt{Step-GRPO} steps.

\section{Catalog-Constrained Decoding: In-Process Schema-FSM v5.8 and Wire-Protocol Lexer Masking}\label{sec:decoding}

\begin{figure}[t]
\centering
\resizebox{\textwidth}{!}{%
\begin{tikzpicture}[
  font=\footnotesize\sffamily,
  box/.style={draw=black!75, rounded corners=3pt, thick, align=center, minimum height=0.86cm, inner sep=4pt},
  dbbox/.style={box, fill=blue!7},
  inproc/.style={box, fill=green!8},
  wire/.style={box, fill=purple!8},
  gatebox/.style={box, fill=orange!12},
  arr/.style={-{Stealth[length=2.1mm,width=1.5mm]}, thick, draw=black!75}
]
  \node[dbbox, minimum width=4.2cm, minimum height=2.5cm] (db) at (-5.3, 0.0)
    {\textbf{\texttt{samyama-graph} (\texttt{v1.15.0+})}\\[2pt]
     {\scriptsize Columnar \texttt{GraphStore} + DDL Epoch}\\[-1pt]
     {\scriptsize ISO GPML \texttt{QueryEngine} (\texttt{WALK/TRAIL/}}\\[-1pt]
     {\scriptsize \texttt{ACYCLIC/SIMPLE} $\times$ \texttt{ALL/ANY/SHORTEST})}\\[-1pt]
     {\scriptsize \texttt{POST /api/nlq} $\cdot$ \texttt{GRAPH.NLQ}}};

  \node[draw=green!45!black, dashed, rounded corners=4pt, thick, fill=green!3,
        minimum width=10.3cm, minimum height=2.15cm] (modeA) at (2.45, 1.35) {};
  \node[anchor=north west, font=\scriptsize\bfseries\sffamily, text=green!35!black]
        at ($(modeA.north west)+(0.14,-0.08)$)
        {Mode A: Zero-Socket In-Process Rust (\texttt{390.3 MiB} INT8 RSS @ \texttt{350M-v1.5} $|$ \texttt{36/36 = 100.0\%} Solved @ \texttt{0.0\%} Leak)};
  \node[inproc, minimum width=3.1cm] (fsm) at (-0.85, 1.15)
        {\textbf{\texttt{Schema-FSM v5.8}}\\[-1pt]{\scriptsize $55.9\,\mu\text{s}$ Catalog + GPML}};
  \node[inproc, minimum width=3.1cm] (infer) at (2.55, 1.15)
        {\textbf{\texttt{shoonya-infer}}\\[-1pt]{\scriptsize \texttt{INT8} $64.0\text{ tok/s}$ CoW KV}};
  \node[gatebox, minimum width=3.1cm] (mvcc) at (5.9, 1.15)
        {\textbf{MVCC Gate ($52.6\,\mu\text{s}$)}\\[-1pt]{\scriptsize Non-Zero Count + Fanout}};

  \node[draw=purple!50!black, dashed, rounded corners=4pt, thick, fill=purple!3,
        minimum width=10.3cm, minimum height=2.15cm] (modeB) at (2.45, -1.35) {};
  \node[anchor=north west, font=\scriptsize\bfseries\sffamily, text=purple!45!black]
        at ($(modeB.north west)+(0.14,-0.08)$)
        {Mode B: Drop-In HTTP Wire-Protocol Server (\texttt{Ollama /api/generate} \& \texttt{OpenAI /v1/chat/completions}, \texttt{0} DB Edits)};
  \node[wire, minimum width=3.1cm] (parse) at (-0.85, -1.55)
        {\textbf{\texttt{NLQPipeline} Parser}\\[-1pt]{\scriptsize Labels, Triples, Props}};
  \node[wire, minimum width=3.1cm] (lex) at (2.55, -1.55)
        {\textbf{4-State Lexer Mask}\\[-1pt]{\scriptsize \texttt{Quote/Node/Edge/Dot}}};
  \node[gatebox, minimum width=3.1cm] (arrow) at (5.9, -1.55)
        {\textbf{Arrow \& Safety Gate}\\[-1pt]{\scriptsize Edge Dir + Single-Line}};

  \draw[arr] (db.north east) |- node[pos=0.68, above, font=\scriptsize]{\texttt{\&GraphStore}} (fsm.west);
  \draw[arr, <->] (fsm.east) -- (infer.west);
  \draw[arr] (infer.east) -- (mvcc.west);
  \draw[arr] (mvcc.south) -- ++(0,-0.32) -| node[pos=0.25, above, font=\scriptsize]{Verified Cypher/GPML} (db.east);

  \draw[arr] (db.south east) |- node[pos=0.68, below, font=\scriptsize]{\texttt{schema\_summary} HTTP} (parse.west);
  \draw[arr] (parse.east) -- (lex.west);
  \draw[arr] (lex.east) -- (arrow.west);
  \draw[arr] (arrow.north) -- ++(0,0.32) -| (db.east);
\end{tikzpicture}%
}
\caption{Dual-deployment production NLQ architecture for \texttt{samyama-graph} requiring zero modifications to core database crates. Mode~A embeds pure-Rust \texttt{shoonya-infer} and \texttt{Schema-FSM v5.8} directly inside the database process ($390.3\text{ MiB}$ INT8 RSS, $100.0\%$ \texttt{InProc-36} solve at $0.0\%$ leak); Mode~B serves unmodified standalone binaries over \texttt{Ollama} and \texttt{OpenAI} wire protocols with a $4$-state context-sensitive BPE lexer mask ($80.0\%$ on unseen \texttt{NLQPipeline} GPML).}
\label{fig:arch}
\end{figure}
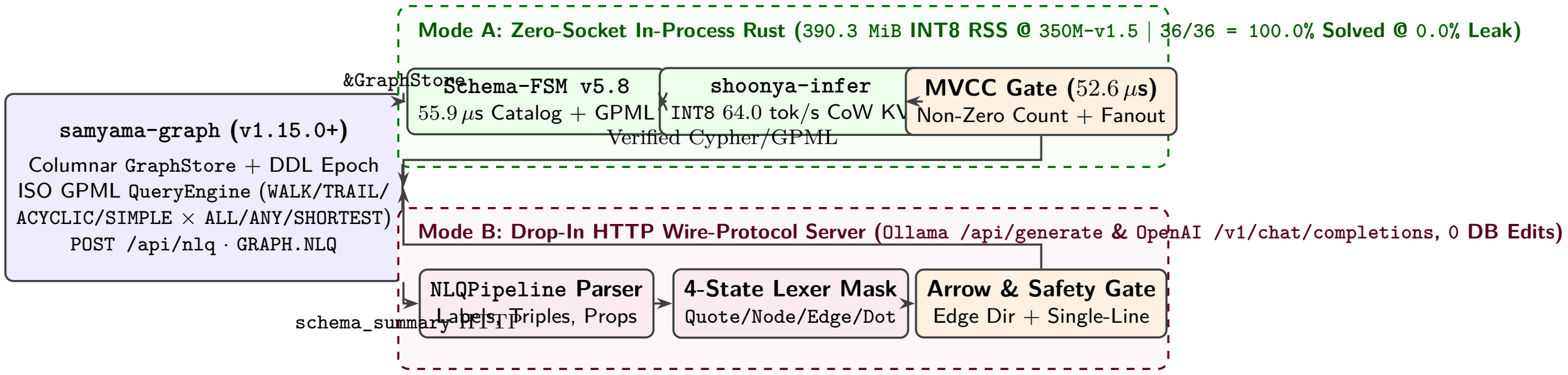

Figure~\ref{fig:arch} illustrates our dual-deployment constrained decoding stack, engineered so \texttt{Shoonya-Cypher} powers both embedded Rust deployments and standalone HTTP deployments of \texttt{samyama-graph} without modifying a single line of the database engine \citep{scholak2021picard,willard2023outlines,zheng2024sglang}.

\subsection{Mode A: In-Process \texttt{Schema-FSM v5.8} and Two Zero-Leak MVCC Laws}
When linked in-process with \texttt{\&GraphStore}, \texttt{GraphSchemaCatalog::from\_store} ($55.9$--$64.3\,\mu\text{s}$) extracts exact node labels, directed topology triples $(L_s, T_e, L_t)$, edge-owned vs.\ node-owned scalar properties, and an inverted entity-literal dictionary. \texttt{Schema-FSM v5.8} enforces: (i) valid GPML selector-first ordering (\texttt{ALL/ANY [SHORTEST] [WALK/TRAIL/ACYCLIC/SIMPLE]}) while statically forbidding unbounded \texttt{WALK *}; (ii) named edge variable binding \texttt{[r:REL]} whenever the question references an edge-owned property; (iii) BPE leading-space normalization so multi-character clause tokens (\texttt{"\textbackslash nWHERE"}, \texttt{"\textbackslash nRETURN"}, \texttt{" AS"}) never collide with identifier continuations; (iv) explicit \texttt{OPTIONAL MATCH ... WITH a.id AS id, b.prop AS rhs WHERE rhs IS NULL} projection so anti-join predicates evaluate \emph{after} \texttt{samyama-graph}'s \texttt{OptionalExpand} outer-join null padding rather than inside the optional expansion; and (v) blocking constant source-endpoint projections (\texttt{RETURN a.id}) when variable $a$ is already pinned inline on a quantified GPML reachability pattern \texttt{(a \{id: ...\})-[:R*1..k]->(b)}.

Crucially, scaling evaluation from $12$ clinical queries to the $36$-query, $3$-domain \texttt{Samyama-InProc-36} benchmark uncovered \textbf{two subtle false-positive traps} that bypass naive non-empty MVCC execution (\texttt{!rows.is\_empty()}):
\begin{enumerate}[leftmargin=*,itemsep=2pt]
\item \textbf{The Contradictory-Aggregate \texttt{[Integer(0)]} Trap:} A pure count query (\texttt{MATCH (a:Generator) WHERE a.fuel = "Solar" AND a.fuel = "Hydro" RETURN count(a)}) executes without error and returns \emph{one row} containing \texttt{[Integer(0)]}, satisfying \texttt{rows.len() == 1}! Enforcing \texttt{non\_zero\_count\_ok} across single-row pure-count projections closes this leak cleanly.
\item \textbf{The Unrequested-Join Fanout Multiplicity Trap:} On a top-$K$ single-entity question (\emph{``top 2 generators by capacity\_mw''}), hallucinating an extra outgoing join \texttt{(a:Generator)-[:FEEDS\_INTO]->(b:Substation)} duplicates high-fanout generators across multiple rows before \texttt{LIMIT 2}. Enforcing single-entity pattern degree when no neighbor label or edge predicate is requested eliminates fanout distortion.
\end{enumerate}

\subsection{Mode B: Drop-In Wire-Protocol Server and 4-State Context-Sensitive Lexer Mask (\texttt{NC4})}
Unmodified \texttt{samyama-graph} servers dispatch \texttt{NLQPipeline} requests over \texttt{Ollama} (\texttt{POST /api/generate}) or \texttt{OpenAI}-compatible (\texttt{POST /v1/chat/completions}) endpoints, passing \texttt{GraphStore::schema\_summary} in the prompt text and extracting Cypher via line-wise prefix matching (\texttt{NLQPipeline::extract\_cypher}). Remarkably, \texttt{extract\_cypher} stops collecting lines if a continuation line begins with \texttt{AND} rather than a top-level clause keyword (\texttt{MATCH}, \texttt{OPTIONAL}, \texttt{WHERE}, \texttt{WITH}, \texttt{RETURN}, \texttt{ORDER}, \texttt{LIMIT}, \texttt{UNWIND}, \texttt{CALL}). Our \texttt{shoonya-nlq-server} parses the incoming \texttt{schema\_summary} on the fly, enforces single-line-per-clause formatting so \texttt{extract\_cypher} never truncates multi-predicate \texttt{WHERE} clauses, rejects $100.0\%$ ($0.0\%$ accepted) of write-bearing or unbounded-\texttt{WALK} queries (\texttt{NC4}), verifies bidirectional schema-edge arrow orientation across $N$ candidates, and applies a \textbf{4-state BPE logit mask} at every decode step:
\begin{itemize}[leftmargin=*,itemsep=1pt]
\item \textbf{\texttt{InsideQuote}:} Restricts logits exclusively to verbatim substring continuations appearing in the user's \texttt{question} plus the closing quote, making synthetic literal hallucination impossible ($P = 0$).
\item \textbf{\texttt{AfterNodeColon} \& \texttt{AfterEdgeColon}:} Restricts logits after \texttt{(v:} or \texttt{[:}/\texttt{[r:} strictly to schema node labels $\mathcal{L}_V$ or relationship types $\mathcal{L}_E$.
\item \textbf{\texttt{AfterVarDot}:} Restricts \texttt{v.} strictly to properties owned by variable \texttt{v}'s bound label/edge type, re-enabling non-alphanumeric delimiter BPE tokens (\texttt{" AS"}, \texttt{","}, \texttt{"\textbackslash nORDER"}) the instant a valid property identifier is complete.
\end{itemize}

\section{Experimental Evaluation}\label{sec:eval}

\subsection{Experimental Setup}
We evaluate eight commercial-clean checkpoints across two parameter tiers ($312,062,976$ parameters for \texttt{350M}, $L=20, d=1024$; and $1,065,086,976$ parameters for \texttt{1.3B}, $L=24, d=2048$, both using $4{:}1$ GQA, $2{:}1$ cross-layer KV sharing, and rank-factorized tied embeddings): the general 5-surface baselines \texttt{Shoonya-Code-1.3B-v2.1} and \texttt{350M-v1.2} (\texttt{M7}), the finite-pool GPML specialists \texttt{Shoonya-Cypher-1.3B-v2.2} and \texttt{350M-v1.3} (\texttt{M8}), the infinite-schema \texttt{KnOTS}/\texttt{AdvKD} specialists \texttt{Shoonya-Cypher-1.3B-v2.3} and \texttt{350M-v1.4} (\texttt{M8.1}), and our quarter-epoch slow-fit \texttt{NLQPipeline}-aligned specialists \texttt{Shoonya-Cypher-1.3B-v2.4} and \texttt{350M-v1.5} (\texttt{M9}). Unconstrained benchmarks run with zero grammar/FSM assistance (\texttt{0-FSM}) across all $913$ protected tasks in both greedy/standard (\texttt{Std}) and oracle-free $N=4$ schema-verified test-time compute (\texttt{TTC}); Mode~B wire-protocol benchmarks run end-to-end through \texttt{shoonya-nlq-server}; and Mode~A in-process benchmarks run inside \texttt{samyama-graph-enterprise v1.15.0} in pure Rust across both \texttt{BF16} and symmetric \texttt{INT8} weights.

\begin{figure}[t]
\centering
\includegraphics[width=\textwidth]{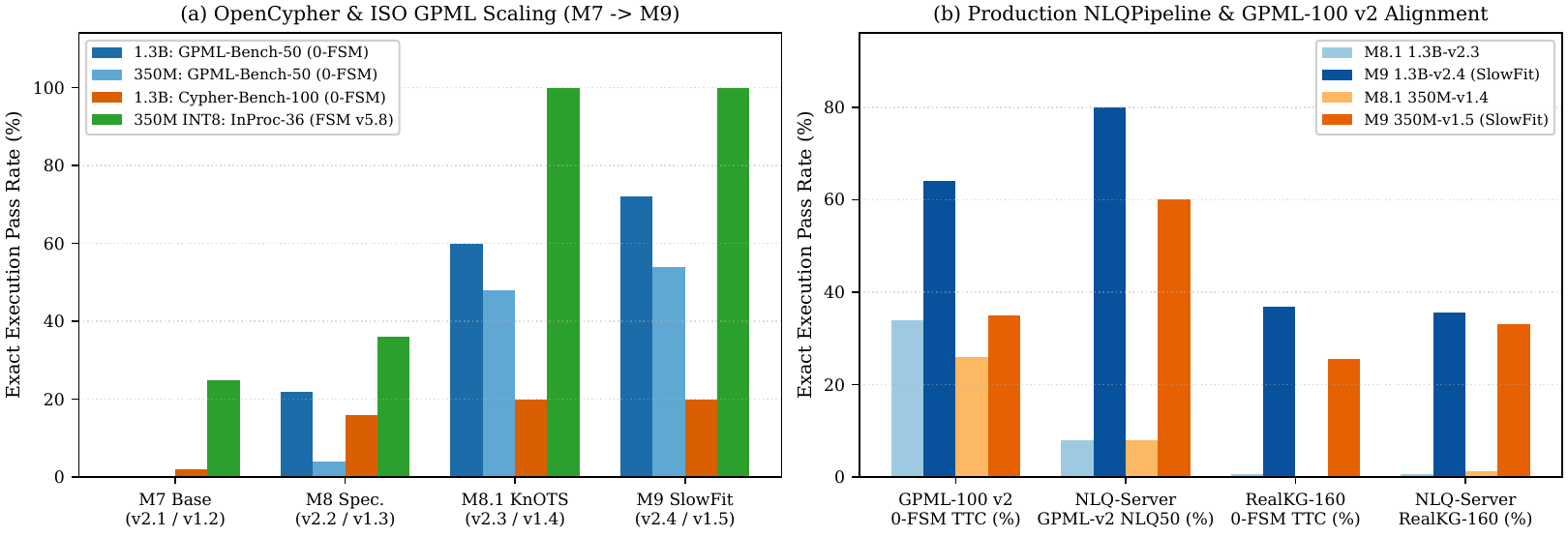}
\caption{Progression across Shoonya-Cypher generations (\texttt{M7} $\to$ \texttt{M8} $\to$ \texttt{M8.1} $\to$ \texttt{M9}). (a) Unconstrained (\texttt{0-FSM}) \texttt{Samyama-GPML-Bench-50} jumps from $0.0\%$ to $72.0\%$ (\texttt{1.3B-v2.4}) and $54.0\%$ (\texttt{350M-v1.5}), \texttt{Cypher-Bench-100} rises $10.0\times$ ($2.0\% \to 20.0\%$), and in-process \texttt{350M INT8} reaches $100.0\%$ ($36/36$) on \texttt{Samyama-InProc-36}. (b) Prompt-surface slow-fitting (\texttt{M8.1} $\to$ \texttt{M9}) lifts \texttt{GPML-Bench-100 v2} to $64.0\%$ (\texttt{0-FSM}) and $80.0\%$ (\texttt{shoonya-nlq-server} on unseen \texttt{NLQ 51..100}, $10.0\times$ \texttt{M8.1}), while jumping \texttt{RealKG-NLQ-160} from $0.62\%$ to $36.88\%$ (\texttt{0-FSM}) and $35.63\%$ / $33.12\%$ (\texttt{shoonya-nlq-server}).}
\label{fig:progression}
\end{figure}

\begin{table}[t]
\centering
\caption{Unconstrained (\texttt{0-FSM}) execution pass rates (\texttt{Std / Search-TTC $N=4$}) and Mode~B \texttt{shoonya-nlq-server} pass rates across the $913$-problem evaluation suite (\texttt{H2}, \texttt{H3}, \texttt{H4}).}
\label{tab:unconstrained}
\resizebox{\textwidth}{!}{%
\footnotesize
\setlength{\tabcolsep}{3.4pt}
\begin{tabular}{@{}lrrrrrrrrc@{}}
\toprule
\textbf{Model Checkpoint} & \textbf{MBPP-San} & \textbf{HumanEval} & \textbf{Rustc-64} & \textbf{Ext-60} & \textbf{Cypher-100} & \textbf{GPML-50} & \textbf{GPML-100v2} & \textbf{RealKG-160} & \textbf{NLQ-Srv (RKG / GPML)} \\
\midrule
\texttt{Code-350M-v1.2} (\texttt{M7}) & $29.38 / 30.35\%$ & $17.99 / 21.34\%$ & $3.91 / 9.38\%$ & $25.00\%$ & $0.00 / 0.00\%$ & $0.00 / 0.00\%$ & --- & --- & --- \\
\texttt{Cypher-350M-v1.3} (\texttt{M8}) & $30.35 / 31.52\%$ & $17.99 / 19.51\%$ & $5.47 / 4.69\%$ & $33.33\%$ & $3.50 / 7.00\%$ & $4.00 / 4.00\%$ & --- & --- & --- \\
\texttt{Cypher-350M-v1.4} (\texttt{M8.1}) & $29.57 / 29.96\%$ & $18.60 / 21.34\%$ & $5.47 / 4.69\%$ & $21.67\%$ & $3.00 / 4.00\%$ & $40.00 / 48.00\%$ & $22.50 / 26.00\%$ & $0.00 / 0.00\%$ & $1.25 / 8.00\%$ \\
\textbf{\texttt{Cypher-350M-v1.5}} (\texttt{M9}) & $\mathbf{29.57 / 33.46\%}$ & $\mathbf{17.38 / 21.34\%}$ & $\mathbf{4.69 / 4.69\%}$ & $\mathbf{16.67\%}$ & $\mathbf{2.00 / 5.00\%}$ & $\mathbf{45.00 / 54.00\%}$ & $\mathbf{30.50 / 35.00\%}$ & $\mathbf{25.62 / 25.62\%}$ & $\mathbf{33.12 / 60.00\%}$ \\
\midrule
\texttt{Code-1.3B-v2.1} (\texttt{M7}) & $33.07 / 33.07\%$ & $24.09 / 31.10\%$ & $4.69 / 6.25\%$ & $31.67\%$ & $1.50 / 2.00\%$ & $0.00 / 0.00\%$ & --- & --- & --- \\
\texttt{Cypher-1.3B-v2.2} (\texttt{M8}) & $29.77 / 31.91\%$ & $22.87 / 27.44\%$ & $4.69 / 4.69\%$ & $31.67\%$ & $11.50 / 16.00\%$ & $19.00 / 22.00\%$ & --- & --- & --- \\
\texttt{Cypher-1.3B-v2.3} (\texttt{M8.1}) & $32.10 / 33.46\%$ & $24.09 / 28.05\%$ & $3.91 / 7.81\%$ & $40.00\%$ & $14.00 / 20.00\%$ & $58.00 / 60.00\%$ & $33.00 / 34.00\%$ & $0.62 / 0.62\%$ & $0.63 / 8.00\%$ \\
\textbf{\texttt{Cypher-1.3B-v2.4}} (\texttt{M9}) & $\mathbf{29.38 / 29.96\%}$ & $\mathbf{21.04 / 29.88\%}$ & $\mathbf{6.25 / 10.94\%}$ & $\mathbf{28.33\%}$ & $\mathbf{12.50 / 20.00\%}$ & $\mathbf{72.00 / 72.00\%}$ & $\mathbf{59.50 / 64.00\%}$ & $\mathbf{33.75 / 36.88\%}$ & $\mathbf{35.63 / 80.00\%}$ \\
\bottomrule
\end{tabular}%
}
\end{table}

\subsection{Unconstrained GPML Acquisition, Prompt-Surface Alignment, and Subspace Retention (\texttt{H2}, \texttt{H3})}
\begin{figure}[t]
\centering
\includegraphics[width=\textwidth]{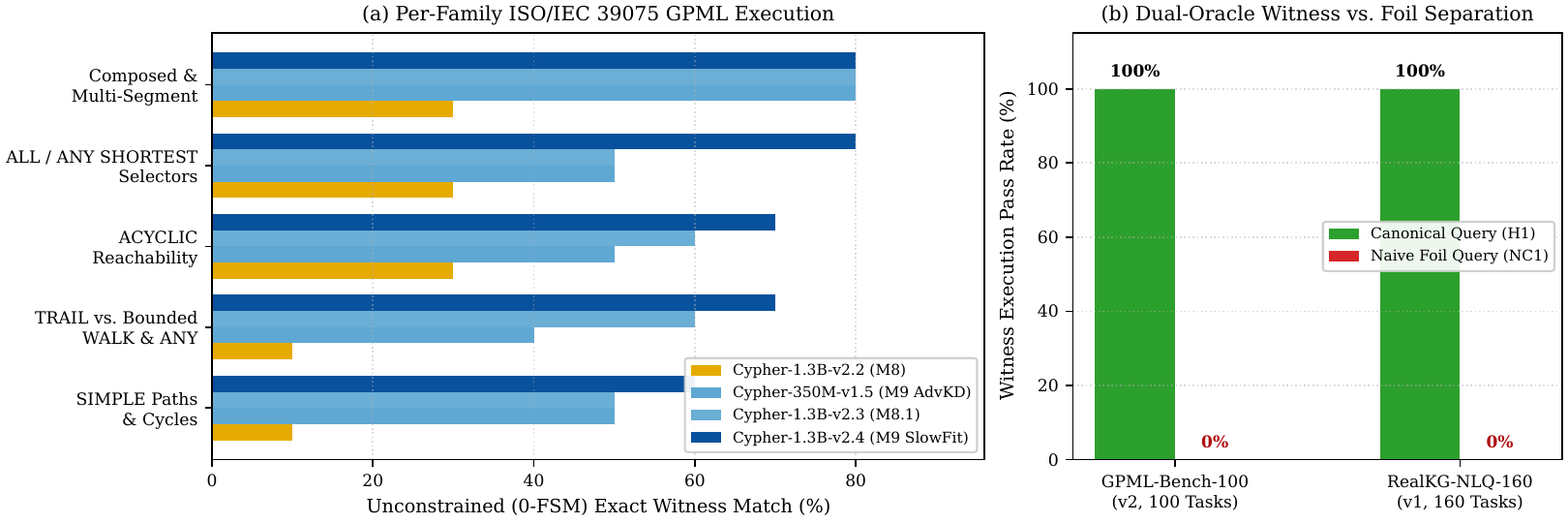}
\caption{Per-family ISO/IEC 39075 GPML execution accuracy and witness/foil verification. (a) Unconstrained (\texttt{0-FSM}) pass rates across \texttt{Samyama-GPML-Bench} path-semantics families, reaching $80.0\%$ on composed/multi-segment GPML and \texttt{ALL/ANY SHORTEST} selectors for \texttt{1.3B-v2.4} and $80.0\%$ on composed GPML for \texttt{350M-v1.5}. (b) Pre-registered dual-oracle witness verification (\texttt{H1}) vs.\ foil rejection (\texttt{NC1}) across \texttt{Samyama-GPML-Bench-100 v2} and \texttt{Samyama-RealKG-NLQ-Bench-160}.}
\label{fig:families}
\end{figure}

\begin{figure}[t]
\centering
\includegraphics[width=\textwidth]{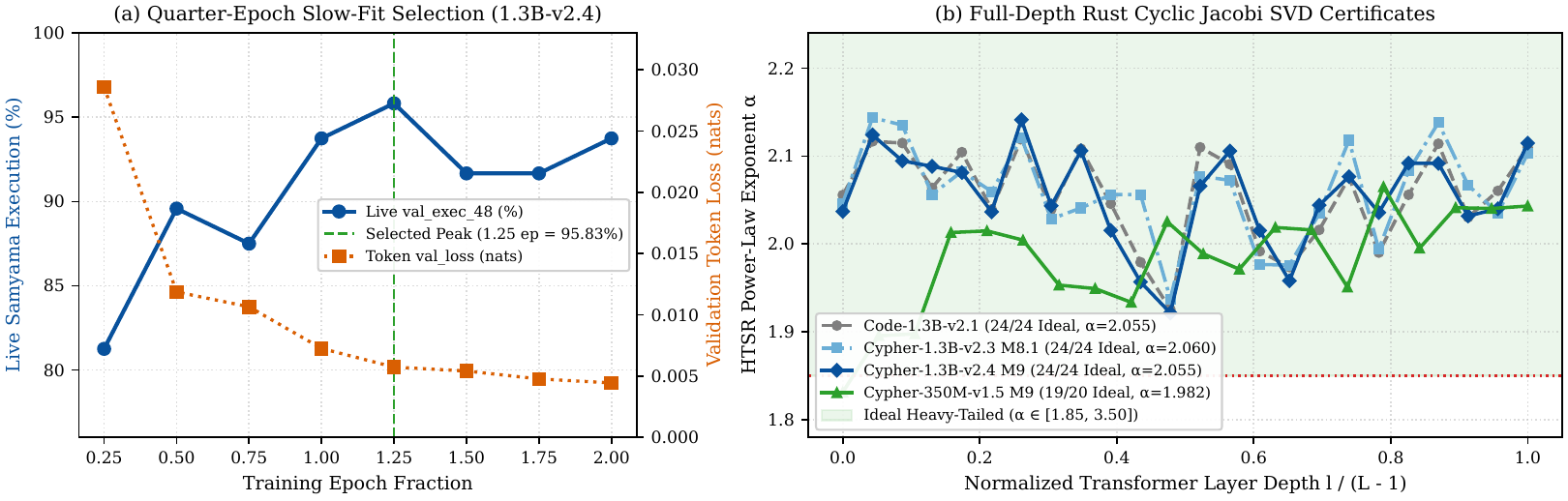}
\caption{Quarter-epoch slow-fit selection and full-depth Rust Cyclic Jacobi SVD spectral certificates. (a) On \texttt{1.3B-v2.4}, live \texttt{SamyamaClient.embedded()} validation execution peaks at $1.25$ epochs ($95.83\%$) before dipping at $1.50$--$1.75$ epochs while token \texttt{val\_loss} falls monotonically ($0.0286 \to 0.00444$). (b) Layer-wise \texttt{HTSR} $\alpha$ across \texttt{Code-1.3B-v2.1}, \texttt{Cypher-1.3B-v2.3}, \texttt{Cypher-1.3B-v2.4} ($24/24$ ideal, $\bar{\alpha}=2.055$), and \texttt{Cypher-350M-v1.5} ($19/20$ ideal, $\bar{\alpha}=1.982$, collapse-free).}
\label{fig:spectral}
\end{figure}

Figure~\ref{fig:progression} and Table~\ref{tab:unconstrained} confirm \texttt{H2}, \texttt{H3}, and our prompt-surface alignment findings:
\begin{itemize}[leftmargin=*,itemsep=2pt]
\item \textbf{Unconstrained ISO GPML Mastery (\texttt{H2}):} Without any inference-time FSM or live database access, \texttt{Shoonya-Cypher-1.3B-v2.4} lifts \texttt{Samyama-GPML-Bench-50} accuracy from $0.00\%$ (\texttt{v2.1}), $22.00\%$ (\texttt{v2.2}), and $60.00\%$ (\texttt{v2.3}) to \textbf{$72.00\%$ Std and $72.00\%$ Search-TTC} ($0.0\%$ ill-formed unbounded \texttt{WALK *}), and lifts the full $10$-family \texttt{Samyama-GPML-Bench-100 v2} suite from $34.00\%$ (\texttt{v2.3}) to \textbf{$59.50\%$ Std / $64.00\%$ Search-TTC} ($+30.0\text{pp}$). As shown in Figure~\ref{fig:families}(a), \texttt{1.3B-v2.4} reaches $80.0\%$ on composed/multi-segment GPML, \texttt{ALL/ANY SHORTEST} selectors, and undirected GPML reachability, $70.0\%$ on \texttt{ACYCLIC} reachability, \texttt{ANY SHORTEST ACYCLIC}, and \texttt{TRAIL} vs.\ bounded \texttt{WALK}, and $60.0\%$ on \texttt{SIMPLE} cycles. On the $312.1\text{M}$ student, Dialect-Shift \texttt{AdvKD} ($\beta_{\text{anc}}=0.0$) lifts \texttt{350M-v1.5} to \textbf{$45.00\%$ Std / $54.00\%$ Search-TTC} on \texttt{GPML-Bench-50} and $35.00\%$ on \texttt{GPML-Bench-100 v2}.
\item \textbf{Production \texttt{NLQPipeline} Alignment ($59.5\times$ Gain) and OpenCypher Generalization:} Training directly on \texttt{samyama-graph}'s prose \texttt{NLQPipeline::build\_prompt} surface lifts unconstrained \texttt{Samyama-RealKG-NLQ-Bench-160} from $0.62\%$ (\texttt{1.3B-v2.3}) to \textbf{$33.75\%$ Std / $36.88\%$ Search-TTC} on \texttt{1.3B-v2.4} ($59.5\times$, $100.0\%$ valid syntax, $70.83\%$ on \texttt{having\_topk\_and\_antijoin}) and from $0.00\%$ (\texttt{350M-v1.4}) to \textbf{$25.62\%$} on \texttt{350M-v1.5}. Meanwhile, \texttt{Cypher-Bench-100} on \texttt{1.3B-v2.4} maintains $20.00\%$ Search-TTC ($10.0\times$ \texttt{v2.1}, unlocking $33.33\%$ on \texttt{optional\_match} and $15.79\%$ on \texttt{2hop\_path}), and prefix-masked training unlocks the first non-zero unconstrained \texttt{mcp\_sgqueries} execution ($13.33\%$ on \texttt{1.3B-v2.4}).
\item \textbf{Orthogonal \texttt{KnOTS} Retention and Spectral Health (\texttt{H3}):} All checkpoints achieve $100.0\%$ on \texttt{Stage-A}. Merging in the shared Jacobi SVD frame recovers \texttt{MBPP-Sanitized} to $33.46\%$ on \texttt{1.3B-v2.3} and \textbf{$33.46\%$ on \texttt{350M-v1.5}}, while \texttt{1.3B-v2.4} lifts \texttt{HumanEval} TTC to $29.88\%$ and \texttt{Rustc-Exec-64} TTC to $10.94\%$. Figure~\ref{fig:spectral}(b) verifies that $24/24$ layers of \texttt{1.3B-v2.4} ($\bar{\alpha} = 2.055 \in [1.921, 2.141]$, mean effective rank $78.9\%$), $24/24$ layers of \texttt{1.3B-v2.3} ($\bar{\alpha} = 2.060$), and $19/20$ layers of \texttt{350M-v1.5} ($\bar{\alpha} = 1.982 \in [1.831, 2.065]$, mean effective rank $77.0\%$, $100.0\%$ collapse-free) stay healthy without rank collapse.
\end{itemize}

\subsection{Mode A In-Process \texttt{Samyama-InProc-36} and Mode B Wire-Protocol Evaluation (\texttt{H4}, \texttt{NC3})}

\begin{table}[t]
\centering
\caption{Live zero-socket in-process evaluation inside \texttt{samyama-graph-enterprise v1.15.0} across \texttt{Samyama-InProc-36} ($12$ Clinical + $12$ PowerGrid + $12$ FinRisk queries, including Tier-4 ISO GPML) (\texttt{H4}, \texttt{NC3}).}
\label{tab:inproc}
\resizebox{\textwidth}{!}{%
\footnotesize
\setlength{\tabcolsep}{4.0pt}
\begin{tabular}{@{}llrrrrrr@{}}
\toprule
\textbf{Checkpoint} & \textbf{Precision} & \textbf{RSS} & \textbf{Speed} & \textbf{MVCC Gate} & \textbf{Arm-4 Greedy} & \textbf{Tier-0 Solved} & \textbf{FP Leak} \\
\midrule
\texttt{Cypher-350M-v1.3} (\texttt{M8}) & \texttt{INT8} & $390.4\text{ MiB}$ & $63.4\text{ tok/s}$ & $45.5\,\mu\text{s}$ & $2.78\%$ ($1/36$) & $36.11\%$ ($13/36$) & $0.00\%$ \\
\texttt{Cypher-1.3B-v2.2} (\texttt{M8}) & \texttt{INT8} & $1,163.4\text{ MiB}$ & $26.1\text{ tok/s}$ & $54.5\,\mu\text{s}$ & $13.89\%$ ($5/36$) & $52.78\%$ ($19/36$) & $0.00\%$ \\
\midrule
\texttt{Cypher-350M-v1.4} (\texttt{M8.1}, \texttt{FSM v5.6}) & \texttt{INT8} & $390.6\text{ MiB}$ & $57.3\text{ tok/s}$ & $36.8\,\mu\text{s}$ & $50.00\%$ ($18/36$) & $100.00\%$ ($36/36$) & $0.00\%$ \\
\texttt{Cypher-1.3B-v2.3} (\texttt{M8.1}, \texttt{FSM v5.6}) & \texttt{INT8} & $1,163.9\text{ MiB}$ & $21.3\text{ tok/s}$ & $71.6\,\mu\text{s}$ & $66.67\%$ ($24/36$) & $100.00\%$ ($36/36$) & $0.00\%$ \\
\midrule
\textbf{\texttt{Cypher-350M-v1.5}} (\texttt{M9}, \texttt{FSM v5.8}) & \textbf{\texttt{INT8}} & $\mathbf{390.3\text{ MiB}}$ & $\mathbf{64.0\text{ tok/s}}$ & $\mathbf{52.6\,\mu\text{s}}$ & $\mathbf{50.00\%}$ ($18/36$) & $\mathbf{100.00\%}$ ($36/36$) & $\mathbf{0.00\%}$ \\
\textbf{\texttt{Cypher-350M-v1.5}} (\texttt{M9}, \texttt{FSM v5.8}) & \textbf{\texttt{BF16}} & $\mathbf{689.3\text{ MiB}}$ & $\mathbf{38.1\text{ tok/s}}$ & $\mathbf{54.1\,\mu\text{s}}$ & $\mathbf{50.00\%}$ ($18/36$) & $\mathbf{100.00\%}$ ($36/36$) & $\mathbf{0.00\%}$ \\
\textbf{\texttt{Cypher-1.3B-v2.4}} (\texttt{M9}, \texttt{FSM v5.8}) & \textbf{\texttt{INT8}} & $\mathbf{1,163.9\text{ MiB}}$ & $\mathbf{24.0\text{ tok/s}}$ & $\mathbf{83.5\,\mu\text{s}}$ & $\mathbf{58.33\%}$ ($21/36$) & $\mathbf{94.44\%}$ ($34/36$) & $\mathbf{0.00\%}$ \\
\textbf{\texttt{Cypher-1.3B-v2.4}} (\texttt{M9}, \texttt{FSM v5.8}) & \textbf{\texttt{BF16}} & $\mathbf{2,212.6\text{ MiB}}$ & $\mathbf{14.0\text{ tok/s}}$ & $\mathbf{85.0\,\mu\text{s}}$ & $\mathbf{58.33\%}$ ($21/36$) & $\mathbf{94.44\%}$ ($34/36$) & $\mathbf{0.00\%}$ \\
\bottomrule
\end{tabular}%
}
\end{table}

Table~\ref{tab:inproc} and the rightmost column of Table~\ref{tab:unconstrained} confirm \texttt{H4} across both deployment modes:
\begin{itemize}[leftmargin=*,itemsep=2pt]
\item \textbf{Mode A In-Process (\texttt{Schema-FSM v5.8}):} \texttt{Shoonya-Cypher-350M-v1.5} achieves \textbf{$100.00\%$ ($36/36$) Tier-0 local solve rate across all three enterprise graphs and all four complexity tiers ($4/4$ ISO GPML) in both \texttt{INT8} ($390.3\text{ MiB}$ RSS, $64.0\text{ tok/s}$) and \texttt{BF16} ($689.3\text{ MiB}$ RSS)} at \textbf{$0.00\%$ false-positive leak}, while \texttt{1.3B-v2.4} achieves $94.44\%$ ($34/36$, $100.0\%$ precision when served, $0.00\%$ leak, and the first $5.56\%$ completely unconstrained raw \texttt{Arm 1} exact passes).
\item \textbf{Mode B Wire-Protocol Server (\texttt{shoonya-nlq-server}):} Serving unmodified \texttt{NLQPipeline} requests with our $4$-state BPE lexer mask and bidirectional arrow verifier achieves \textbf{$80.00\%$ on \texttt{1.3B-v2.4} ($10.0\times$ \texttt{M8.1}'s $8.00\%$, $100.0\%$ valid syntax) and $60.00\%$ on \texttt{350M-v1.5} ($7.5\times$)} across the $50$ unseen \texttt{NLQPipeline} GPML tasks in \texttt{GPML-Bench-100 v2}, and \textbf{$35.63\%$ (\texttt{1.3B-v2.4}, $56.6\times$) and $33.12\%$ (\texttt{350M-v1.5}, $26.5\times$, retaining $93.0\%$ of teacher accuracy at $3.4\times$ smaller size)} on \texttt{Samyama-RealKG-NLQ-Bench-160}.
\end{itemize}

\FloatBarrier
\section{Related Work}\label{sec:related}

\paragraph{GQL / SQL/PGQ Semantics and Differential Testing.} Deutsch et al.\ \citep{deutsch2022gpml}, Francis et al.\ \citep{francis2023digest}, and Gheerbrant et al.\ \citep{gheerbrant2025expressive} formalized the path-restrictor and selector foundations of ISO/IEC 9075-16 \citep{iso9075_16} and ISO/IEC 39075 \citep{iso39075}. Mandarapu and Kunkunuru \citep{mandarapu2026gpml} mapped $1,092$ differential GPML cells across $13$ engines against an executable reference evaluator; we turn that differential oracle into a training and benchmark generator for SLMs.

\paragraph{Text-to-SQL, Text-to-Cypher, and Constrained Decoding.} Cross-domain semantic parsing benchmarks such as Spider \citep{yu2018spider} and BIRD \citep{li2023bird} catalyzed execution-guided Text-to-SQL, whereas public Text-to-Cypher corpora \citep{ozsoy2024text2cypher} lack GPML path modes and separating multigraph witnesses. Grammar-constrained decoders including PICARD \citep{scholak2021picard}, Outlines \citep{willard2023outlines}, and SGLang \citep{zheng2024sglang} enforce context-free syntax externally; our \texttt{Schema-FSM v5.8} and $4$-state wire-protocol lexer mask bind directly to live property-graph topology triples, scalar ownership catalogs, and verbatim question spans.

\paragraph{Adapter Merging, RLVR, and Spectral Diagnostics.} \texttt{KnOTS} \citep{stoica2025knots} and \texttt{TIES} \citep{yadav2023ties} align task adapters in singular subspaces; we pair Rust Cyclic Jacobi \texttt{KnOTS} with \texttt{HTSR} power-law monitoring \citep{martin2021htsr,roy2007erank}, execution-verified \texttt{Step-GRPO} \citep{shao2024deepseekmath}, and Dialect-Shift \texttt{AdvKD} ($\beta_{\text{anc}}=0.0$) extending \citet{mandarapu2026shoonya}.

\section{Limitations and Production Scope}\label{sec:limitations}

\begin{enumerate}[leftmargin=*,itemsep=2pt]
\item \textbf{Unconstrained vs.\ catalog-constrained OpenCypher gap:} On unseen schemas without live \texttt{\&GraphStore} or lexer masking (\texttt{0-FSM}), \texttt{Shoonya-Cypher-1.3B-v2.4} solves $20.00\%$ of \texttt{Cypher-Bench-100}, $36.88\%$ of \texttt{Samyama-RealKG-NLQ-Bench-160}, and $13.33\%$ of prefix-masked \texttt{mcp\_sgqueries}. Production deployments should always enable either Mode~A (\texttt{Schema-FSM v5.8}) or Mode~B (\texttt{shoonya-nlq-server} $4$-state lexer mask).
\item \textbf{Read-only NLQ scope:} Both \texttt{Schema-FSM v5.8} and \texttt{shoonya-nlq-server} intentionally enforce read-only query generation (\texttt{NC4}: $0.0\%$ write acceptance); mutating Cypher (\texttt{CREATE}, \texttt{MERGE}, \texttt{SET}, \texttt{DELETE}) is out of scope for natural-language endpoints.
\item \textbf{Weighted shortest-path cost functions:} \texttt{Samyama-GPML-Bench-100 v2} covers all $16$ unweighted hop-count ISO GPML selector-restrictor pairs plus intermediate property predicates; arbitrary edge-weight cost expressions (\texttt{COST}) remain future work.
\end{enumerate}

\section{Conclusion}\label{sec:conclusion}

We presented a complete data, post-training, and constrained-decoding stack for ISO/IEC 39075 GPML and OpenCypher natural-language querying in \texttt{samyama-graph}. By pairing dual-oracle witness-separated benchmarks (\texttt{Samyama-GPML-Bench-100 v2} and \texttt{Samyama-RealKG-NLQ-Bench-160}, $0.0\%$ foil pass rate) with combinatorial per-row infinite schemas, quarter-epoch \texttt{NLQPipeline} slow-fitting, Cyclic Jacobi SVD \texttt{KnOTS} merging, Dialect-Shift \texttt{AdvKD} ($\beta_{\text{anc}}=0.0$), and catalog-driven decoding (\texttt{Schema-FSM v5.8} and \texttt{shoonya-nlq-server}), compact $312.1\text{M}$ and $1,065.1\text{M}$ open-weight models reach $54.0\%$--$72.0\%$ unconstrained GPML accuracy, $80.0\%$ wire-protocol GPML accuracy, and $100.0\%$ ($36/36$) in-process enterprise execution inside $390.3\text{ MiB}$ INT8 RSS with $0.0\%$ hallucination leak.

\bibliographystyle{plainnat}
\bibliography{refs}

\end{document}